\documentclass[10pt,conference]{IEEEtran} 
\usepackage{hyperref}
\usepackage{amsmath}
\usepackage{adjustbox}

\def\BibTeX{{\rm B\kern-.05em{\sc i\kern-.025em b}\kern-.08em
    T\kern-.1667em\lower.7ex\hbox{E}\kern-.125emX}}
    \makeatletter
\newcommand{\linebreakand}{%
  \end{@IEEEauthorhalign}
  \hfill\mbox{}\par
  \mbox{}\hfill\begin{@IEEEauthorhalign}
}

\ifCLASSINFOpdf
\else
\fi
\usepackage{graphicx}
\usepackage{dcolumn}

\newcolumntype{d}[1]{D{.}{.}{#1}}

\begin{document}
%
\title{The Introduction of README and CONTRIBUTING Files in Open Source Software Development}




%
\author{\IEEEauthorblockN{Matthew Gaughan\IEEEauthorrefmark{1},
Kaylea Champion\IEEEauthorrefmark{2},
Sohyeon Hwang\IEEEauthorrefmark{3}, 
Aaron Shaw\IEEEauthorrefmark{1}} 

\IEEEauthorblockA{\IEEEauthorrefmark{1}Northwestern University, Evanston, Illinois, USA\\
 Email: gaughan@u.northwestern.edu}

\IEEEauthorblockA{\IEEEauthorrefmark{2}University of Washington, Seattle, Washington, USA}

\IEEEauthorblockA{\IEEEauthorrefmark{3}Princeton University, Princeton, New Jersey, USA}}

%

\maketitle

\begin{abstract}
\texttt{README} and \texttt{CONTRIBUTING} files can serve as the first point of contact for potential contributors to free/libre and open source software (FLOSS) projects. Prominent open source software organizations such as Mozilla, GitHub, and the Linux Foundation advocate that projects provide community-focused and process-oriented documentation early to foster recruitment and activity. In this paper we investigate the introduction of these documents in FLOSS projects, including whether early documentation conforms to these recommendations or explains subsequent activity. We use a novel dataset of FLOSS projects packaged by the Debian GNU/Linux distribution and conduct a quantitative analysis to examine \texttt{README} (n=4226) and \texttt{CONTRIBUTING} (n=714) files when they are first published into projects' repositories. We find that projects create minimal \texttt{README}s proactively, but often publish \texttt{CONTRIBUTING} files following an influx of contributions. The initial versions of these files rarely focus on community development, instead containing descriptions of project procedure for library usage or code contribution.  The findings suggest that FLOSS projects do not create documentation with community-building in mind, but rather favor brevity and standardized instructions. 
\end{abstract}


%
\IEEEpeerreviewmaketitle

\section{Introduction}
In the early stages of FLOSS project development, the sustainability of the project depends on the recruitment of experienced, committed contributors \cite{xiao_how_2023}. Open source guides propose that public documentation---such as \texttt{README} and \texttt{CONTRIBUTING} files---provides a key entry point, recruitment resource, and mechanism for establishing project goals, norms, and procedures. For example, GitHub notes that \texttt{README} files help maintainers communicate expectations for their projects and manage contributions.\footnote{\href{https://docs.github.com/en/repositories/managing-your-repositorys-settings-and-features/customizing-your-repository/about-readmes}{https://docs.github.com/en/repositories/managing-your-repositorys-settings-and-features/customizing-your-repository/about-readmes}} Prominent community leaders also argue that well-constructed documents facilitate scaling.\footnote{\href{https://github.com/hackergrrl/art-of-readme}{https://github.com/hackergrrl/art-of-readme}} The Mozilla Open Leadership Training site recommends that projects provide explicit contributor guidelines (usually, \texttt{CONTRIBUTING} files) ``that explain how new contributors can help out on your project,''\footnote{\href{https://mozilla.github.io/open-leadership-training-series/articles/building-communities-of-contributors/write-contributor-guidelines/}{https://mozilla.github.io/open-leadership-training-series/articles/building-communities-of-contributors/write-contributor-guidelines/}} and the Cloud Native Computing Foundation claims that such guidelines improve contribution quality.\footnote{\href{https://contribute.cncf.io/maintainers/templates/contributing/}{https://contribute.cncf.io/maintainers/templates/contributing/}} These recommendations from leaders in the FLOSS ecosystem assume that documentation files catalyze growth and contribution activity by attracting and building contributor community.

However, despite prevailing wisdom that ``good'' documentation contributes to successful, sustainable FLOSS projects, empirical study of these documents and their relationship to subsequent project growth and contributions remains scarce. Widespread belief in the value of documentation may lead projects to create files in a ritualistic, hollow way simply because doing so has become the norm. 
Alternatively, documentation may only arrive after projects have already become established, emerging as a consequence rather than a cause of growth or success along other dimensions. Further, claims that specific qualities of the content of documentation shape subsequent project activity merit empirical investigation. Understanding the significance of documentation more broadly entails examining both the introduction of documentation as well as whether documentation attributes help explain project outcomes. 

In this study, we evaluate the introduction of \texttt{README} and \texttt{CONTRIBUTING} files---two of the most common types of documentation about community procedures---in a large heterogeneous sample of FLOSS projects with diverse approaches to documentation. From an initial population of over 21,000 projects packaged in the Debian GNU/Linux operating system distribution, we identify 4,226 upstream projects with \texttt{README} files and 714 projects with \texttt{CONTRIBUTING} files. We parse the contribution logs for these projects and construct an original panel dataset to examine when they introduce these documents, document characteristics at the point of introduction, and the correlations of those initial characteristics with variations in subsequent project activity. We conduct a mixed methods documentation analysis and multilevel longitudinal analysis to describe how projects introduce documentation. While previous research has analyzed project documentation files, we believe that this is one of the first empirical studies evaluating the initial publication of \texttt{README} and \texttt{CONTRIBUTING} files in open source software.

Our results indicate that while \texttt{README} documents are standard early-stage additions, \texttt{CONTRIBUTING} documents are often published later, following a flurry of contributions. The initial versions of most \texttt{README} documents are very short, less than 160 words and requiring only a few seconds to read. Though the majority of \texttt{CONTRIBUTING} documents are still relatively short (less than 250 words) they tend to be longer than \texttt{README} files. In terms of content, we find that initial versions of the files lack community-related language and are more technical in focus. Regarding claims that introducing documentation will catalyze growth via subsequent contributor activity, we find no evidence to support such causality, though we do find associations between \texttt{CONTRIBUTING} files focusing on code style and subsequent contribution activity.

%
Critically, we find that document publication often follows periods of increased contribution activity and precedes periods of diminished activity. Thus, our study indicates that projects neither develop \texttt{README} and \texttt{CONTRIBUTING} files consistent with prevailing wisdom nor in a way consistent with a causal relationship to project growth or contribution activity. 



\section{Background \& Related Works}

\subsection{Documentation in FLOSS Development Processes}
Documentation serves as a key means by which FLOSS projects communicate with prospective contributors about the features, goals, and organization of a project. 
The aims and processes that projects pursue, such as formal code reviews, have important consequences for the quality of software that results \cite{bosu_peer_2013}. Projects must balance tensions between reducing barriers to contribution (facilitating more activity from more volunteers) and maintaining the quality of the code that gets contributed \cite{de_stefano_impacts_2022, meneely_secure_2009, meneely_strengthening_2010}. Trustworthy and transparent projects better recruit and retain volunteers \cite{chakraborti_we_2023, gallivan_striking_2001, guizani_attracting_2022}. %
As a result, recommendations that projects create friendly, readable, public documentation in order to facilitate newcomer contributions are widespread \cite{abdalla_template_2016}. Previous work has found that even out of date or poorly structured documentation can support continued engagement with the project \cite{forward_relevance_2002} and may help newcomers overcome barriers to entry \cite{forte_defining_2013}. 

\texttt{README} and \texttt{CONTRIBUTING} files have become the overwhelmingly common methods of communicating how a project can be accessed, used, and contributed to \cite{oakley_hackergrrlart--readme_2024}. 
Recommendations that projects create and use \texttt{README} and \texttt{CONTRIBUTING} files early in their life cycles to recruit and retain contributors have become commonplace throughout FLOSS ecosystems and other peer production systems. We provide specific examples and elaborate details of such recommendations below, but they usually entail two key assumptions: first, that project leaders want to grow and sustain a community of contributors; and second, that the files will support sustained growth by offering accessible, well-crafted information about how to get involved, establishing and demonstrating a welcoming community culture. Such claims gain support from prior work illustrating how clear communication of peer production project goals and ways of working may get large, collaborative endeavors off the ground in the first place \cite{hill_essays_2013}.


Several recent studies cast doubt on these assumptions. Founders of peer production projects do not always seek to grow large communities \cite{foote_starting_2017}. Likewise, people do not always decide to join communities due to their size or growth \cite{hwang_why_2021}. Among software projects, \texttt{README} files contain more content about the usage of the project than assisting prospective contributors \cite{prana_categorizing_2019}, even though developers evaluate \texttt{README} files when deciding where to submit pull requests \cite{qiu_signals_2019}. While \texttt{CONTRIBUTING} documents often address would-be developers, they rarely focus on the barriers encountered by contributors \cite{fronchetti_contributing_2023}; when they do, contributors rarely adhere to documents' prescribed processes~\cite{elazhary_as_2019}. The apparent gap between the prevailing wisdom about the value of documentation files (i.e., guiding contributions toward building a robust community of contributors) and the empirical analyses of files raises questions about how projects create \texttt{README} and \texttt{CONTRIBUTING} files with regards to the broader prevailing wisdom.

The initial versions of \texttt{README} and \texttt{CONTRIBUTING} files in FLOSS projects offers unique insight into how projects approach such documentation. Prior work has conducted cross-section analyses of the latest versions of documentation files \cite{prana_categorizing_2019, tan_how_2024, fronchetti_contributing_2023}. However, the content and purpose of documentation may change over a project's life cycle. The latest versions cannot explain how or when projects introduced such files, which offers signals as to why projects adopt documentation. The point of introduction of \texttt{README} and \texttt{CONTRIBUTING} files is also significant because the first-version files likely have an out-sized role in shaping later document versions and subsequent organizational processes: prior literature in political science \cite{pierson_path_2000} and management \cite{Sydow_Schreyogg_Koch_2009} illustrates how early actions determine later organizational decisions via path dependence. 

We focus on \texttt{README} and \texttt{CONTRIBUTING} files because these two kinds of documents are well-established, widely used, and the subject of explicit recommendations from both prior research and leading FLOSS organizations. We elaborate on these points below as we describe the two in greater depth to motivate our specific research questions and analyses. Additionally, because \texttt{README} and \texttt{CONTRIBUTING} files are so widespread, they enable analysis at a scale that supports inferences across a large and heterogeneous sample of FLOSS projects. We further discuss this when we introduce our research setting, sample, and analysis approach in the Methods section.

\subsection{\texttt{README} Documents}

\texttt{README} files have been used in software projects since at least the mid-1970s to centralize project information for users and developers.\footnote{An example from a PDP 10 in 1974 may be found at: \href{http://pdp-10.trailing-edge.com/decus_20tap3_198111/01/decus/20-0079/readme.txt.html}{http://pdp-10.trailing-edge.com/decus\_20tap3\_198111/01/decus/20-0079/readme.txt.html}} 
A recent study found nearly ubiquitous adoption of \texttt{README} files across both continuously-active and deprecated projects \cite{coelho_why_2017}. As the ``project's first impression,'' recommendations suggest that \texttt{README} file touch on a wide range of topics, from routing help requests to testing.\footnote{For example, see \href{https://github.com/cfpb/open-source-project-template/blob/main/README.md}{https://github.com/cfpb/open-source-project-template/blob/main/README.md} and \href{https://gitlab.com/tgdp/templates/-/blob/main/readme/guide-readme.md}{https://gitlab.com/tgdp/templates/-/blob/main/readme/guide-readme.md}} 



Open source guides suggest that \texttt{README} documents should be structured, parsimonious, and detailed, providing enough documentation for the reader to adeptly make use of the software and to draw in prospective contributors.\footnote{See: \href{https://www.makeareadme.com/}{https://www.makeareadme.com/}} Prior empirical research finds that projects seem to follow at least some of these recommendations. Cross-sectional analysis has found that though documents rarely follow the structures recommended by GitHub, project contributors eventually develop lengthy \texttt{README} files \cite{liu_how_2022}. Many \texttt{README} files contain functional information regarding how to use the project and what the project accomplishes, but contain little information on community or contributions \cite{prana_categorizing_2019}. In addition, many \texttt{README} files provide sufficient detail to support automated project builds \cite{hassan_mining_2017}; and  documentation of engineering procedure that contains detailed usage examples facilitates higher developer productivity \cite{sohan_study_2017}. In some cases, \texttt{README} documents support contributor recruitment and retention \cite{qiu_signals_2019} and identify project contribution policies \cite{tan_how_2024}. However, while \texttt{README} files may evolve into useful documents that support multiple goals in a software development community, prior work does not explore the contents of \texttt{README} files when they are first created. 


\subsection{\texttt{CONTRIBUTING} Guidelines}

In contrast to the recommendations that \texttt{README} documents act as comprehensive guides to projects, open source guides recommend using \texttt{CONTRIBUTING} documents as focused tools to recruit and structure community contributions.\footnote{For example, contrast \href{https://github.com/cfpb/open-source-project-template/blob/main/CONTRIBUTING.md}{https://github.com/cfpb/open-source-project-template/blob/main/CONTRIBUTING.md} with \href{https://github.com/cfpb/open-source-project-template/blob/main/README.md}{https://github.com/cfpb/open-source-project-template/blob/main/README.md}} 
CONTRIBUTING documents define the rules and processes of how contributions---primarily code commits---are made to the project.
\cite{gousios_exploratory_2014, hassan_studying_2006}. Previous studies have identified useful traits of contributing guidelines for community development, such as the convenient presentation of information~\cite{vincent_deeper_2021}, description of recruitment contributor recruitment processes \cite{fronchetti_contributing_2023}, and description of community mentorship programs \cite{balali_newcomers_2018}. However, although these guidelines may make first impressions for prospective project contributors \cite{qiu_signals_2019}, previous research has found that \texttt{CONTRIBUTING} files often neither address common issues of project contributors nor are adhered to by those contributors \cite{fronchetti_contributing_2023, elazhary_as_2019}. As with \texttt{README} files, we focus on \texttt{CONTRIBUTING} files' introduction to projects because the timing and content of initial versions can offer insight into how projects approach and use such files. 


Finally, for FLOSS projects, a primary governance concern is ensuring that the diverse motivations of project contributors align with volunteering time and energy towards a given project \cite{di_tullio_governance_2013}. The recruitment and management of code contributions are a primary topic of FLOSS project rules \cite{crowston_coordination_2005}. The governance of development processes is crucial to the successful recruitment of volunteer labor: just as the management of code commits can engage contributors, so too can projects demotivate further contributions through governance that is inequitable or overly restrictive \cite{alami_how_2020, steinmacher_almost_2018}. Recommendations for the development of \texttt{README} and \texttt{CONTRIBUTING} files often imply that the introduction of a document will cause the project to attract additional contributions.\footnote{\textit{e.g.}.~, see href{https://gitlab.com/tgdp/templates/-/blob/main/contributing-guide/guide-contributing-guide.md}{https://gitlab.com/tgdp/templates/-/blob/main/contributing-guide/guide-contributing-guide.md}} By analyzing contribution activity that follows the introduction of these files, we offer an initial evaluation of such claims. 


\section{Study Design}

We structure our analysis around five research questions to understand the introduction of governance documents: 
\begin{itemize}
    \item \textbf{RQ1: }\textit{When are \texttt{README} documents published to a project?}
    \item \textbf{RQ2:} \textit{What do maintainers include in the first versions of their \texttt{README} documents?}
    \item \textbf{RQ3: }\textit{When are \texttt{CONTRIBUTING} documents published to a project?}
    \item \textbf{RQ4:} \textit{What do maintainers include in the first versions of their \texttt{CONTRIBUTING} guidelines?}
    \item \textbf{RQ5:} \textit{How do characteristics of initial \texttt{README} and \texttt{CONTRIBUTING} files relate to subsequent activity?}
\end{itemize}

Our research questions motivate a large-scale analysis comparing documentation and contribution activity across a heterogeneous sample of FLOSS projects. To pursue this, we created a novel dataset of activity and document data that we collected from FLOSS projects packaged in the Debian GNU/Linux distribution.\footnote{Our data and code are available on the Harvard Dataverse: [\href{https://doi.org/10.7910/DVN/LEFZKR}{https://doi.org/10.7910/DVN/LEFZKR}]} These projects' inclusion in the Debian package distribution indicates that they are all widely used libraries that underpin broader FLOSS ecosystems. We investigate our research questions through three analyses: a longitudinal analysis of document introduction, a descriptive analysis of initial document characteristics, and an analysis of how initial document characteristics relate to later project contribution activity.

\subsection{Setting and Data Collection}

The Debian GNU/Linux ecosystem has a substantial history as a setting for academic research into software engineering communities and their governance \cite{champion_underproduction_2021, claes_historical_2015, ververs_influences_2011, krafft_how_2016}. The Debian distribution is also a source of packages for Ubuntu and several other GNU/Linux distributions. By studying the upstream projects of packages included in Debian, we are studying a sample of FLOSS projects that are developed by heterogeneous communities and organizations and that are important to the broader software ecosystem. Debian also does not specify rigid project governance or documentation requirements, resulting in diverse documentation practices. This is in contrast to other potential samples of projects (used in other prior work) drawn from a given umbrella foundation or ecosystem (e.g. Apache or Eclipse), which must adhere to specific, explicitly defined organizational norms in their development and maintenance. 

We restricted our data set to packages hosted on Debian's Salsa platform and with an upstream project that used git as the primary version control system, regardless of where their upstream repository was hosted (e.g. GitHub, GitLab). The Debian project recommends that all packages within their distribution maintain information regarding upstream project location in the package files themselves, in the \texttt{upstream/package/metadata} directory.\footnote{\href{https://dep-team.pages.debian.net/deps/dep12/}{https://dep-team.pages.debian.net/deps/dep12/} (Archived:\href{https://perma.cc/9J2H-S8VP}{https://perma.cc/9J2H-S8VP})} Nevertheless, compliance with this recommendation remains low across bundled packages, and only one third of eligible Debian packages specified their upstream repository. Moreover we experienced some initial data loss as certain packages' upstream projects required developer credentials, resulting in access limitations. We drew from an initial list of the 21,902 Debian packages studied in Champion and Hill \cite{champion_underproduction_2021} to define an intermediary data set consisting of 5,092 upstream projects.  

Our initial sampling identified projects from the intermediary data set that contained \texttt{README} or \texttt{CONTRIBUTING} files in the root directory of their main branches on March 16, 2024. We collected project commit history with the GitPython library.\footnote{\href{https://gitpython.readthedocs.io/en/stable/intro.html}{https://gitpython.readthedocs.io/en/stable/intro.html}} From our intermediary data set, we were unable to clone 706 projects due to SSH git cloning errors.
Of the 4,386 projects that we were able to access, we identified 85 that lacked either a \texttt{README} or \texttt{CONTRIBUTING} file in their root directory. Using GitPython, we then found the first commit in each project's history which contained either a \texttt{README} or \texttt{CONTRIBUTING} document, irrespective of file type. This step generated a small amount of errors in downloading the document at commit time.
Our final dataset consisted of \texttt{README} (n=4226) and \texttt{CONTRIBUTING} (n=714) files from 4247 projects (693 projects were represented in both datasets) that all use the git VCS, are widely used through Debian packaging, and have accessible upstream project source code. 

For both datasets, we collected weekly commit data for the six months before and after document introduction, as well as new contributor data for the two months before and after document introduction. Preliminary analysis showed large amounts of documentation commit activity in the hours following file introduction. To capture a more stable first version of the document, we collected the contents of the \texttt{README} and \texttt{CONTRIBUTING} file versions that existed one week after file initialization. 

\subsection{Modeling Document Introduction}
\label{sec: mlm-methods}
\begin{figure}[t!]
    \centering
        \includegraphics[width=\linewidth]{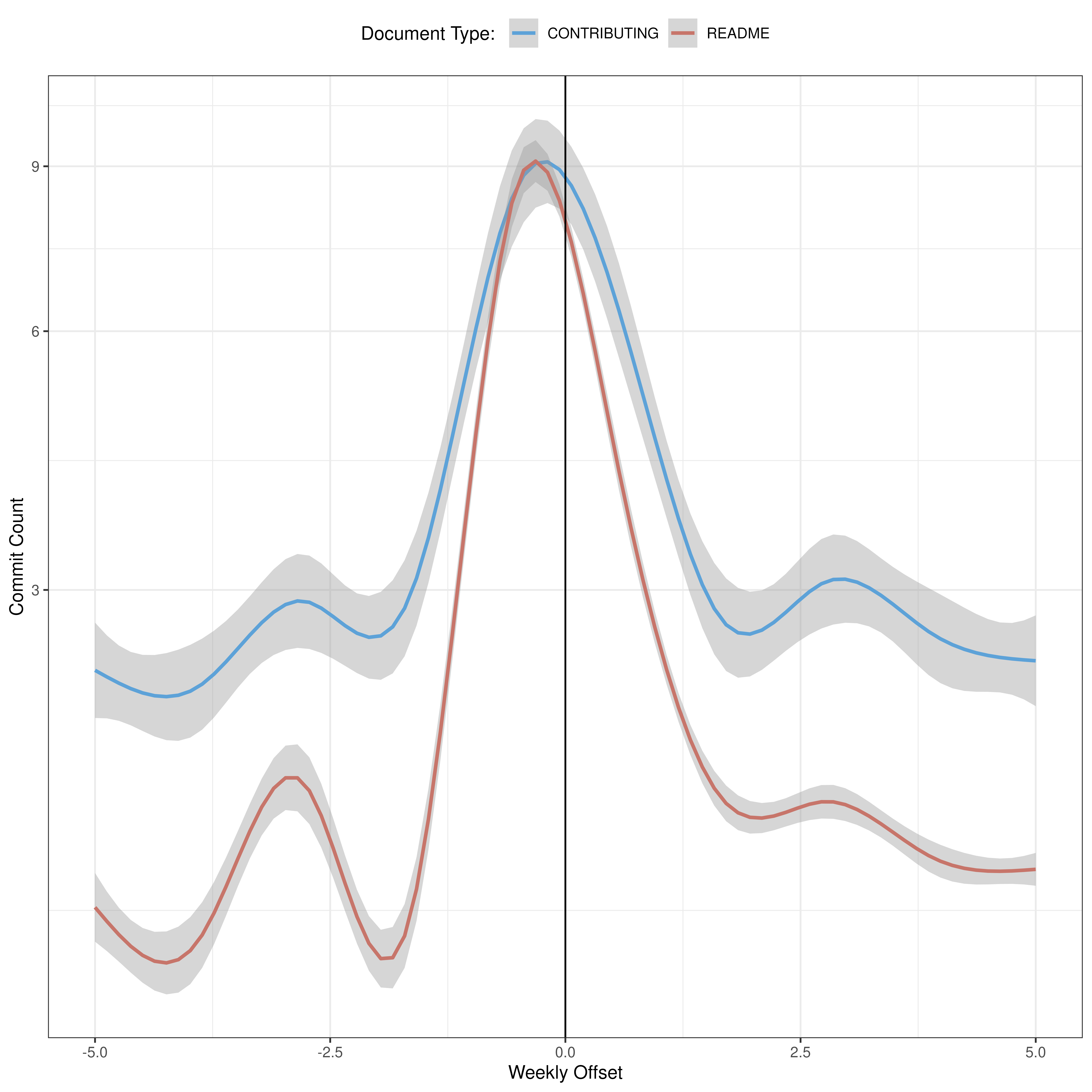}
        \caption[Plot of average (log-transformed) weekly contribution counts over time around the point of document introduction (weeks offset from document publication date) for \texttt{README} (red) and \texttt{CONTRIBUTING} (blue) files. The Y-axis has been scaled to real count values.]{Plot of average (log-transformed) contribution counts over time around the point of document introduction (weeks offset from document publication date) for \texttt{README} (red) and \texttt{CONTRIBUTING} (blue) files.The Y-axis has been scaled to real count values. \footnotemark}\label{fig:draft_averages}
    \end{figure}

We construct multilevel longitudinal models to further understand the timing of document introduction and document attributes at the point of introduction (\textbf{RQ1}, \textbf{RQ3}), as well as the relationship of document attributes to subsequent contribution activity (\textbf{RQ5}). To understand when documents are typically introduced into projects, we evaluated distributions of weekly project contribution counts and age around the point at which documents were published. The average distribution of weekly project contributions surrounding document publication is shown in Figure \ref{fig:draft_averages}. Then, we cast our data as weekly contribution counts nested within projects. This results in a hierarchical data structure that requires a mixed effects time series approach \cite{steele_multilevel_2008} in order to account for the dependencies between errors as well as the variance. The mean and variance for weekly commit counts (per project) were 4.31 and 207.76 for projects in our \texttt{README} data set. For projects in our \texttt{CONTRIBUTING} data set, the mean and variance weekly commit counts were 8.37 and 349.06. To account for the shape of the data, we logarithmically transformed our weekly commit counts and used the \texttt{lme4} package in R to fit a negative binomial regression model~\cite{bates_fitting_2015}. 

\footnotetext{Figure color palettes have been selected to support readability \cite{crameri_scientific_2023}.}

\subsubsection{Multilevel Time Series Model Specifications}
To characterize patterns of project activity around the introduction of the documents, we adopt a modeling strategy derived from Regression Discontinuity (RD) methods \cite{murnane_methods_2011}.\footnote{RD methods are most often used to identify causal effects. However, the assumptions necessary to support causal inference in an RD framework are not satisfied in our setting, where the timing of document introduction could have many causes related to the underlying trend in contribution activity. See \cite{murnane_methods_2011} for more on RD methods.} We identified the optimal bandwidth of 10 weeks for our time series analysis through using the Imbens-Kalyalaraman test through the \texttt{rdd} package \cite{imbens_optimal_2009}.\footnote{\href{https://cran.r-project.org/web/packages/rdd/rdd.pdf}{https://cran.r-project.org/web/packages/rdd/rdd.pdf}} The regression formula for the model of project contribution activity as a function of document publication appears in Equation \ref{eq:readme_rdd}, where for each project $j$, the log transformed commit count ($Y$) for each week ($i$) is a function of the intercept ($\beta_{0j}$), the weekly ($W$) growth rate of contributions ($\beta_{1j}$), the impact ($\beta_{2j}$) of document introduction ($D$), document publication impact ($\beta_{3j}$) on the weekly growth rate ($I:W$) and project age ($A_j$). The formula is the same for the modeling of both \texttt{README} and \texttt{CONTRIBUTING} data. While project age remains fixed, the other variables are unique to each weekly data point in our two and a half month window. We evaluated models' autocorrelation function (ACF) plots and variance inflation factors (VIF) to identify possible violations in our model. For both data sets, the ACF plots revealed autocorrelation in small lag values, with slight autocorrelation over larger lag values; our data show clusters of commit activity over a handful of weeks. Models' VIF scores showed multi-colinearity (VIF values above 5) between the $W$ and $D:W$ variables; given both variables' usage of the same weekly index variable, this is expected and not a threat to model validity. 
\begin{equation}
\centering
            Y_{ij} = \beta_{0j} + \beta_{1j}(W) + \beta_{2j}(D) + \beta_{3j}(D:W) + A_j + r_{ij}
\label{eq:readme_rdd} 
\end{equation}
Each $\beta_{ij}$ can be decomposed into its constituent parts of the true effect $\gamma_{ij}$ and error $u_{ij}$, as shown in Equation \ref{eq:beta_distillation}.   
\begin{equation}
\centering
        \beta_{ij} = \gamma_{ij} + u_{ij}
\label{eq:beta_distillation} 
\end{equation}

\subsection{Descriptive Document Analysis}
\label{sec: ta-methods}
To understand document contents at the point of introduction (\textbf{RQ2}, \textbf{RQ4}), we performed computational grounded theory \cite{nelson_computational_2020} to study the first versions of \texttt{README} and \texttt{CONTRIBUTING} files. Given that prevailing sentiment (as articulated in FLOSS community guides) supposes that reader-friendly documents are useful for recruiting contributor activity, we calculated readability metrics for document text. To study the subject matter of documents, we used Latent-Dirichlet Allocation (LDA) topic models. 

We selected three readability metrics to evaluate document form: the Flesch reading ease score, a well-established general setting readability metric \cite{flesch_new_1948}; the linsear write formula, developed for the evaluation of technical documents \cite{klare_assessing_1974}; and the McAlpine EFLAW scoring system \cite{mcalpine_global_1997}, which evaluates the readability of English-language documents for readers whose primary language is not English. These widely used readability metrics provide a general assessment of document characteristics. We used the Python \texttt{textstat} library to calculate all three metrics.\footnote{\href{https://github.com/textstat/textstat}{https://github.com/textstat/textstat}}

To evaluate documents' subject matter, we tuned an LDA model implemented with the Gensim topic modeling library\cite{rehurek_software_2010}. For document stopwords, we used the Natural Language Toolkit English stopwords extended with the 20 most common terms across documents\footnote{These are: ["http", "com", "www", "org", "file", "code", "time", "software", "use", "user", "set", "line", "run", "source", "github", "lineno", "python", "php", "ruby", "api"].} \cite{bird_natural_2009} and stripped any presentation styling (e.g. Markdown, HTML). We used 
Greene et al.'s \cite{greene_how_2014} approach to select LDA topic counts through term-centric stability analysis; producing optimal topic counts of 9 for \texttt{README} documents and 5 for \texttt{CONTRIBUTING} documents. For other LDA model parameters (e.g. learning decay, batch size), we implemented a grid search to optimize other variables for model interpretability.

We labeled LDA topics with natural language descriptors through inductive thematic coding \cite{creswell_research_2009}. Three coauthors independently coded themes from modeled topics and reached consensus on topic descriptors. From these analyses we report topic labels, top 10 keywords from each topic, along with the 25th percentile, median, and 75th percentile of intra-document topic coefficient distributions.

\subsection{Modeling Document Characteristics and Project Outcomes}
Using the results from our descriptive analysis of process documents as well as from our multilevel time series models, we test the relationships between document characteristics and project outcomes such as new contributor activity and commit activity in the eight weeks following \texttt{README} and \texttt{CONTRIBUTING} file publication through negative binomial regression models and descriptive analysis of document groupings. 

\subsubsection{Grouping by Coefficient Estimates}
    \begin{figure}[t!]
    \centering
        \includegraphics[width=\linewidth]{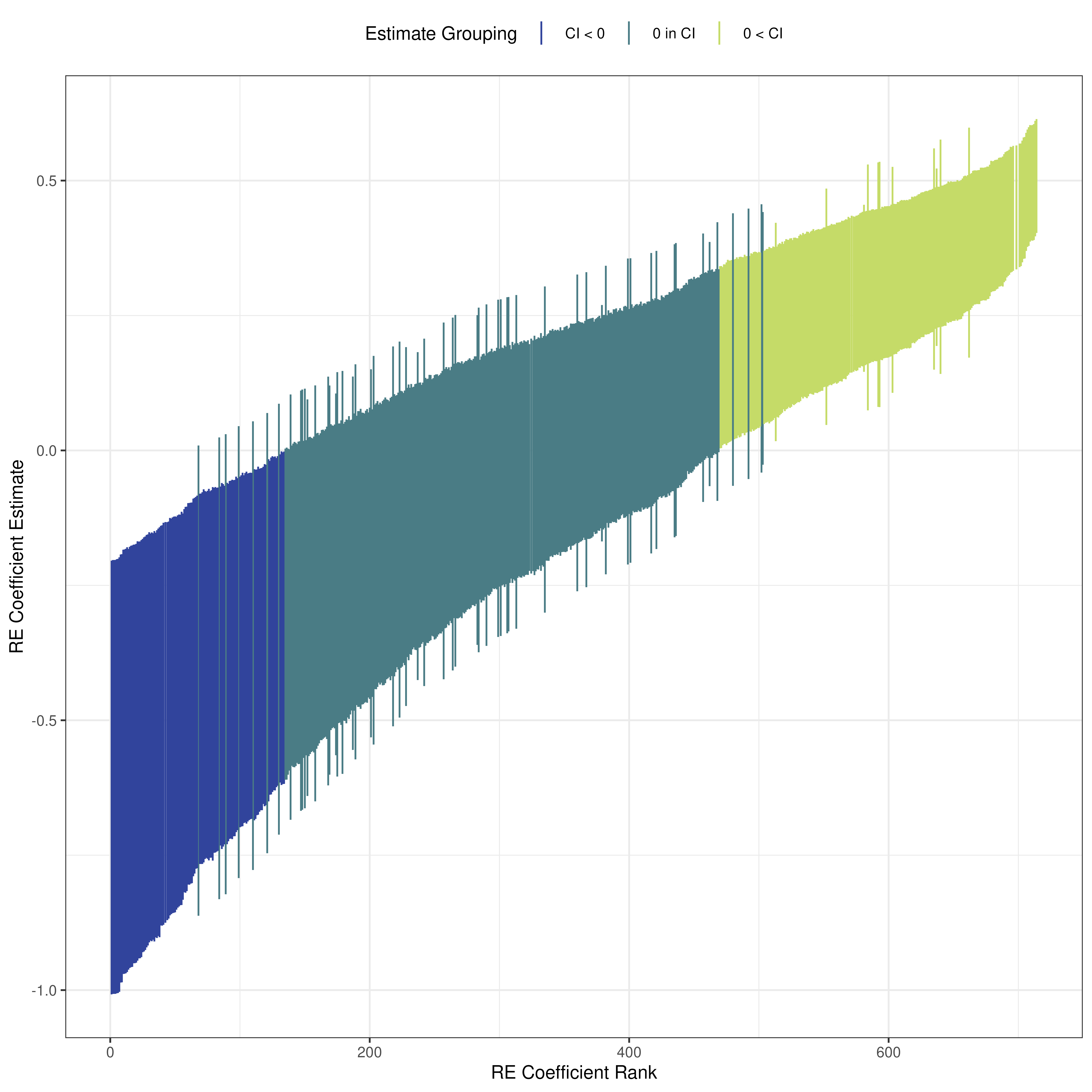}
        \caption{Plot of project-level fitted random effects coefficients and estimated 95\% confidence intervals (Y-axis) sorted by coefficient rank (X-axis) from the model of project activity as a function of \texttt{CONTRIBUTING} file introduction. Colors correspond to groupings of projects based on whether the 95\% CI for the random effects coefficient estimate is less than (dark blue), overlapping with (teal), or greater than (lime) zero.} \label{fig:blup_grouping}
    \end{figure}

Mixed-effects models provide point estimates of the coefficient values for each project's random effects; we follow prior work in zoology and ecology and use these coefficients to identify project-level relationships \cite{robinson_that_1991}. Following prior recommendations for caution in using these coefficient estimates \cite{hadfield_misuse_2010, houslay_avoiding_2017}, we use the confidence intervals of our random effect coefficients from our multilevel time series models to categorize projects according to whether they experience activity decline, stability, or growth. We do this by calculating the coefficient estimates for the interaction effect of the document introduction on the weekly growth rate of commit activity. This random effect indicates the relative weekly change in contributions from before and after documents are introduced; it shows how projects sustain week-over-week contribution growth across the introduction of a given file. By using the 95\% confidence intervals of these estimates, we can specifically evaluate the characteristics of projects that have positive post-introduction growth changes ($CI > 0$), negative post-introduction growth changes ($CI < 0$) and null post-introduction growth changes ($0 \in CI$). This categorization is illustrated for our \texttt{CONTRIBUTING} document projects in Figure \ref{fig:blup_grouping}. The singular fit of both our mixed effects models indicate low variance across random effects coefficients. As such, the majority of projects in both data sets are categorized as experiencing null post-introduction growth changes ($0 \in CI$).

\subsubsection{Modeling Project Activity as a Function of Document Characteristics}

To further investigate the relationships between document contents and project outcomes (\textbf{RQ5}), we model the association between topic distributions with the total sums of commits in the eight weeks following document publication. To estimate these relationships, we again use a negative binomial model, as our outcome (project activity) is a count variable with a leftward skew. For fitting purposes, we removed our model intercept. Thus, we interpret our topic coefficients as the direct effects of a given topic on zero commits following document introduction. Given sparse data for new contributors, we did not find any significant relationships between document traits and contributor recruitment.

We evaluated the relationships between project outcomes ($Y$) and the distribution of our LDA model topics within a project's document. For \texttt{CONTRIBUTING} files, the model formula is shown in Equation \ref{eq:contributing_topic_outcome} where $TC1$ - $TC5$ represent the distribution of the corresponding topic (All topic numbers correspond to those provided in Table \ref{tab:lda-grouped-topics}). We used a similar formula for \texttt{README} files, where $TR1$ -$TR9$ represented the distributions of corresponding topics. 

\begin{equation}
\centering
    Y = \beta_{1}(TC1) + \beta_{2}(TC2) + \beta_{3}(TC3) + \beta_{4}(TC4) + \beta_{5}(TC5) + r
\label{eq:contributing_topic_outcome}
\end{equation}

\section{Results}

\subsection{Modeling Activity Around Document Introduction}


\subsubsection{When do projects usually publish \texttt{README} and \texttt{CONTRIBUTING} files?}

\begin{table}[t!]
\caption{Results for two multilevel negative binomial longitudinal models regressing project activity on project age, time, and the introduction of \texttt{README} (first column) and \texttt{CONTRIBUTING} (second column) files. For each variable in both models, we report fitted coefficient values as well as corresponding 95\% confidence intervals (in brackets).}
\setlength{\tabcolsep}{0.4\tabcolsep}
\centering
\begin{tabular}{l r r}
\hline
 & \texttt{README} &\texttt{CONTRIBUTING}\\
\hline
(Intercept)                                           & $0.193^{*}$         & $0.472^{*}$         \\
                                                     & $ [ 0.130;  0.257]$ & $ [ 0.394;  0.550]$ \\
Introduction                                        & $0.473^{*}$         & $0.250^{*}$         \\
                                                     & $ [ 0.409;  0.538]$ & $ [ 0.172;  0.329]$ \\
Week (Time)                                           & $0.273^{*}$         & $0.168^{*}$         \\
                                                     & $ [ 0.246;  0.299]$ & $ [ 0.144;  0.191]$ \\
Project Age                                          & $-0.032^{*}$        & $-0.010$            \\
                                                     & $ [-0.051; -0.013]$ & $ [-0.053;  0.033]$ \\
Introduction:Week                                     & $-0.684^{*}$        & $-0.407^{*}$        \\
                                                      & $ [-0.714; -0.653]$ & $ [-0.443; -0.371]$ \\
\hline
AIC                                                 & $74892.750$         & $20808.134$         \\
BIC                                                  & $75026.103$         & $20919.005$         \\
Log Likelihood                                      & $-37430.375$        & $-10388.067$        \\
Num. obs.                                            & $30778$ \phantom{.000}             & $7551$\phantom{.000}              \\
Num. groups: Project                    & $4226$\phantom{.000}              & $714$\phantom{.000}               \\
\hline
\multicolumn{3}{l}{\scriptsize{$^*$ Null hypothesis value outside the confidence interval.}}
\end{tabular}
\label{tab:contribution changes}
\end{table}

Our results show that \texttt{README} and \texttt{CONTRIBUTING} files are introduced at different times in project life cycles. 
The median time to \texttt{README} file publication was the on the same day of a project's initial commit, but there is a wide range of when projects publish, with a standard deviation over three years. In contrast, \texttt{CONTRIBUTING} files were introduced far later in project lifespans: the median time to \texttt{CONTRIBUTING} file publication (1806 days) was over four years into project development. 

Multilevel time series models of project activity around the point of document introduction suggest that both \texttt{README} and \texttt{CONTRIBUTING} documents are published around local maxima in contribution activity. For both types of files, project contribution activity declines after their introduction; however, subsequent decreases in commit activity may be due to when the documents are published in the project's life cycle (e.g., right before a release) rather than any effect of the files themselves. Table \ref{tab:contribution changes} shows regression model results for both document types. We summarize the results separately here.

\subsubsection{\texttt{README} Introduction}

For \texttt{README} files, the model results in the first column of Table \ref{tab:contribution changes} indicate that following file publication, weekly commit activity declines (the negative coefficient on the interaction of \textit{Introduction:Week}). When we exponentially transform our model values, our results show that the prototypical \texttt{README} publication follows a flurry of initial commits, file publication is then subsequently followed by an average decline of over six commits every week (\textbf{RQ1}). 

\subsubsection{\texttt{CONTRIBUTING} introduction}

Similarly, for \texttt{CONTRIBUTING} files (the second column of Table \ref{tab:contribution changes}), we observe an uptick in activity during the three weeks prior to introduction. Though activity continues to increase in the week of document introduction (the positive coefficient for \textit{Introduction},) we find a continued decline in contribution activity in the subsequent month (the negative coefficient on the interaction of \textit{Introduction:Week}). Our results show that a few years into project development, contributors typically publish \texttt{CONTRIBUTING} guides following a flurry of activity (\textbf{RQ3}), at which point contribution activity diminishes. 

We can put these results into more concrete terms by describing the model fitted values corresponding to contribution activity for a prototypical (mean activity level) project. Five weeks prior to \texttt{CONTRIBUTING} file publication, contributions to the project are steady week over week. Then, projects typically see contribution growth of 3.90 commits per week in the three weeks prior to document publication. In the week of document publication, projects experience a further increase in contributing activity by about 2.61 commits. Commit activity then tapers at a rate of 3.49 commits per week. 

\subsection{Descriptive Analysis of Document Contents}

    \begin{figure}[t!]
    \centering
        \includegraphics[width=\linewidth]{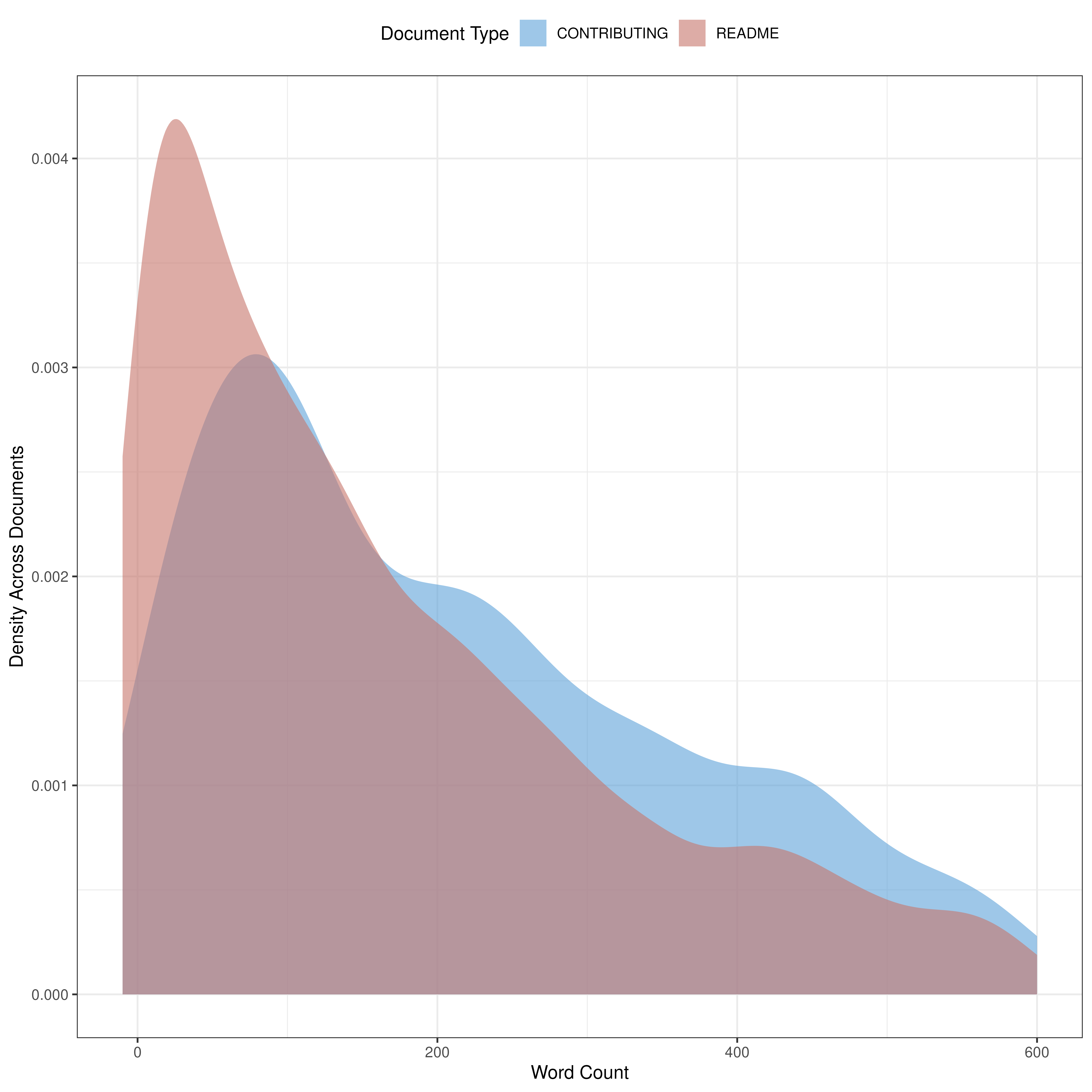}
        \caption{ Plot of the kernel density of document word counts for first-version \texttt{README} (red) and \texttt{CONTRIBUTING} (blue) files.} \label{fig:document_size}
    \end{figure}
\subsubsection{Document Readability and Length}
Our quantitative analysis of initial \texttt{README} and \texttt{CONTRIBUTING} files\footnote{As noted earlier in Study Design, we examine the version one week after file initialization to capture a more stable initial version of the document} found that both document types had similar readability and length, though differed on subject matter. The initial versions of both files shared a similar average reading level around a high-school level of difficulty. \texttt{README} files in our data set had median scores of 16.50 for McAlpine EFLAW, 7.14 for Linsear Write, and 50.86 for Flesch Reading Ease. \texttt{CONTRIBUTING} files in our data set had median scores of 18.15 for McAlpine EFLAW, 7.38 for Linsear Write, and 53.10 for Flesch Reading Ease.

For both types of process documents, initial files were fairly short, the median reading time for \texttt{README} files was 14.79s while the median reading time for \texttt{CONTRIBUTING} files was 19.73s. However, more files in our \texttt{README} data set were empty than files in our \texttt{CONTRIBUTING} data set. Figure \ref{fig:document_size} plots the density of the distribution of word count for \texttt{README} and \texttt{CONTRIBUTING} documents. Both distributions exhibit a substantial left-skew with the bulk of each massed below 200 words. Indeed, the plot shows that most \texttt{README} documents are initially published with little to no content at all (fewer than 100 words). 

\begin{table*}[t]
  \centering
    \caption{Manually labeled topics from our fitted LDA model (left column), the top 10 keywords for each topic (second-left column), as well as the interquartile range for intra-document topic distributions for each topic (rightmost columns). The median coefficient distribution values are bolded. }
  \begin{tabular}{l l r r r }
  Topic (Manually Labeled) & Top 10 Keywords & 25\% & 50 \% & 75\%\\
  \hline
    \textbf{\texttt{README}} &   \\
       R1. Usage (graphics) & image, data, key, file, color, option, support, format, default, mode&  0.001 & \textbf{0.004} & 0.090\\
       R2. Usage (architecture) &   data, test, library, object, implementation, support, packet, used, byte, class & 0.001 & \textbf{0.006} & 0.117\\
       R3. Legal &  license, copyright, perl, gnu, free, version, module, public, general, warranty&  0.001 & \textbf{0.006} & 0.108\\
       R4. Code snippets & test, value, function, return, method, class, string, type, object, example & 0.001 & \textbf{0.004} & 0.102 \\
       R5. Configuration instructions & http, git, server, install, client, request, test, version, project, command& 0.001 & \textbf{0.028} & 0.201\\
       R6. Usage (parsing)& json, node, require, string, parser, var, object, parse, function, font &  0.001 & \textbf{0.004} & 0.072\\
       R7. Usage (command-line) &  command, output, option, process, make, program, script, tool, file, linux & 0.001 & \textbf{0.005} & 0.111\\
       R8. Usage (structured text) &  table, html, tag, text, django, xml, example, path, template, default& 0.001 & \textbf{0.004} & 0.081\\
       R9. Installation instructions &   install, make, build, library, version, directory, file, package, window, project& 0.002 & \textbf{0.148}  & 0.491\\
    \textbf{\texttt{CONTRIBUTING}}&   \\
    C1. General contribution instructions& http, project, git, bug, scipy, contributing, request, pull, issue, contribute
 &  0.001 & \textbf{0.002} & 0.022\\
    C2. Code style guide &  build, function, style, make, file, command, test, used, option, variable
 & 0.001 & \textbf{0.005} & 0.115\\
    C3. Contributing procedure & issue, request, pull, change, bug, feature, branch, git, project, make
  & 0.010 & \textbf{0.542} & 0.823\\
    C4. Contributor agreement &  license, contribution, patch, project, submit, test, open, sign, agreement, & 0.001 & \textbf{0.004} & 0.079\\
    C5. Build instructions &    test, git, make, install, version, doc, http, change, release, commit& 0.002 & \textbf{0.034}& 0.315
  \end{tabular}
  \label{tab:lda-topics}
\end{table*}

\subsubsection{Document Topics}

Our LDA topic models help illustrate the subject matter that project contributors include in the initial versions of both file types. Topic descriptions, prominent keywords, and average intra-document topic distributions are shown in Table \ref{tab:lda-topics}. Similar to the cross-sectional findings of Prana et al. \cite{prana_categorizing_2019}, we find that while many first versions of \texttt{README} documents are sparse, the average \texttt{README} document focuses primarily on technical installation and usage instructions with little attention to community development (\textbf{RQ2}). Shown in $TR1$, $TR2$, $TR6$, $TR7$, and $TR8$, given the prevalence of usage instructions in \texttt{README} files, the LDA model differentiated usage instructions by library functions. Moreover, while \texttt{CONTRIBUTING} documents include a lot of information about the technical processes of building a development environment ($TC5$) and the organizational processes of submitting a code change ($TC3$,) we do not find topics about building community (\textbf{RQ4}). 

\subsection{Document Characteristics and Contribution Activity}


    \begin{figure}[t!]
    \centering
        \includegraphics[width=\linewidth]{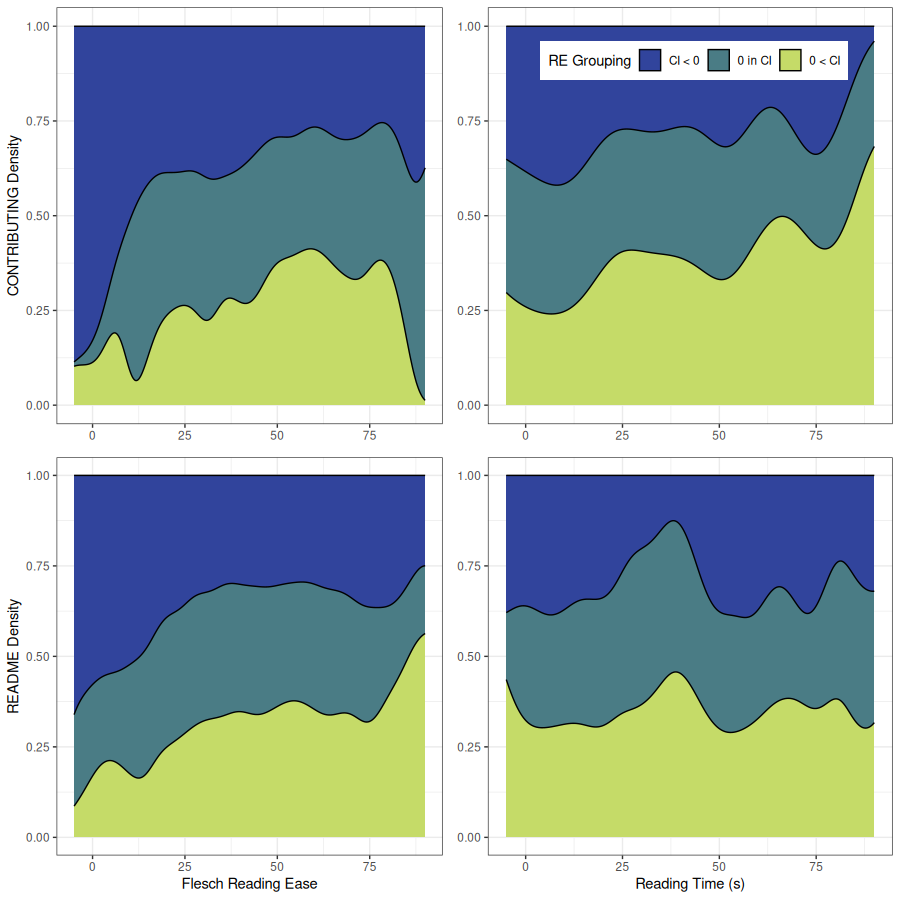}
        \caption{ Plot of the kernel densities for Flesch Reading Ease (left-column) and reading time (right-column) metrics for \texttt{README} (bottom row) and \texttt{CONTRIBUTING} (top row) documents. Colors correspond to groupings of projects based on whether the 95\% CI for the random effects coefficient estimate is less than (dark blue), overlapping with (teal), or greater than (lime) zero.} 
        \label{fig:readability_metrics}
    \end{figure}

For both document types, Figure \ref{fig:readability_metrics} shows the kernel density of readability and reading time grouped by random effects coefficient estimates from our multilevel time series models. Substantial variations in readability or file reading time produces divergent density distributions, which are mapped to the area of each color in the plots. Across coefficient groupings, documents were largely similar in form, with most documents short in length and at a high school to college reading level. However, the Flesch Reading Ease density plot for both files show that projects which published simple documents tended to experience post-publication declines in commit activity. 

Table \ref{tab:lda-grouped-topics} presents the manually labeled LDA model topics with mean intra-document topic distributions grouped by projects' random effects coefficient estimates as previously introduced. Each column shows the prototypical topic distribution of the introduced file of projects that experienced relative post-introduction contribution decline ($CI<0$), stability ($CI \in 0$), or increase ($CI > 0$). Projects that experienced a relative increase in contribution activity ($CI > 0$) following \texttt{README} introduction had 
\texttt{README}s focused more on installation instructions ($TR9$). Projects that experienced increased contribution activity following \texttt{CONTRIBUTING} introduction had \texttt{CONTRIBUTING} documents focused on code style guides ($TC2$) and less information about contributor agreements ($TC4$).

\begin{table}[t]
  \centering
    \caption{Manually labeled topics from the fitted LDA topic models (left column) with the mean intra-document topic distribution grouped by random effects coefficient groupings introduced in Figure \ref{fig:blup_grouping} (right columns).}
  \begin{tabular}{l  r r r}
  Topic (Manually Labeled) & $CI<0$ & $0 \in CI$ & $CI>0$  \\
  \hline
    \textbf{\texttt{README}} &   \\
       1. Usage (graphics) &0.072 & 0.073& 0.073 \\
       2. Usage (architecture) & 0.117& 0.094& 0.111\\
       3. Legal & 0.094 & 0.102 & 0.076\\
       4. Code snippets & 0.056& 0.084 & 0.050\\
       5. Configuration instructions & 0.148& 0.144 & 0.130\\
       6. Usage (parsing) & 0.069& 0.076 & 0.059 \\
       7. Usage (command-line) & 0.124 & 0.093 & 0.093\\
       8. Usage (structured text) & 0.057 & 0.074 & 0.067\\
       9. Installation instructions & 0.262 & 0.259 & 0.339\\
    \textbf{\texttt{CONTRIBUTING}}&   \\
    1. General contribution instructions &  0.090& 0.112 & 0.094 \\
    2. Code style guide &  0.052 &  0.099 & 0.126\\
    3. Contributing procedure & 0.501& 0.451 & 0.535 \\
    4. Contributor agreement & 0.184  & 0.115 & 0.073 \\
    5. Build instructions & 0.173  & 0.222 & 0.172 \\
  \end{tabular}
  \label{tab:lda-grouped-topics}
\end{table}

Fitting negative binomial models on our data, we found significant relationships between document subject matter and subsequent project activity. Table \ref{tab:topics-outcomes} shows the results of our models for both \texttt{README} and \texttt{CONTRIBUTING} data sets.

We found that projects with \texttt{README} files that contain more content pertaining to configuration ($TR5$) and installation ($TR9$) instructions experience more contribution activity in the weeks following document publication than those that do not. Similarly, when \texttt{CONTRIBUTING} files focus on code style guidelines ($TC2$), projects experience more subsequent contributions. Given the variety of possible antecedents of contribution activity, these models do not support causal interpretation. Rather, these results indicate how relative emphasis on certain topics at the point of document introduction is related to higher amounts of contribution activity in the weeks that follow.

\begin{table}
  \centering
    \caption{Fitted negative binomial models regressing project activity (new commits in the eight weeks following document publication) on manually-labeled topics from LDA topic models fitted on \texttt{README} and \texttt{CONTRIBUTING} files. We report 95\% confidence intervals in brackets.}
\begin{tabular}{l r r}
\hline
Topic (Manually Labeled) & \texttt{README} & \texttt{CONTRIBUTING} \\
\hline
\textbf{\texttt{README}} \\
1. Usage (graphics)     & $1.107^{*}$    &   \\
               & $ [0.993; 1.221]$ & \\
2. Usage (architecture)     & $1.221^{*}$      & \\
               & $ [1.130; 1.312]$ & \\
3. Legal       & $0.950^{*}$    &   \\
               & $ [0.863; 1.036]$ & \\
4. Code snippets        & $1.000^{*}$     &  \\
               & $ [0.888; 1.112]$  &\\
5. Configuration instructions        & $1.211^{*}$     &  \\
               & $ [1.141; 1.281]$  &\\
6. Usage (parsing)       & $1.069^{*}$     &  \\
               & $ [0.962; 1.176]$ & \\
7. Usage (command-line)      & $1.096^{*}$      & \\
               & $ [0.998; 1.193]$ & \\
8. Usage (structured text)        & $1.094^{*}$      & \\
               & $ [0.983; 1.205]$  &\\
9. Installation instructions        & $1.154^{*}$      & \\
               & $ [1.109; 1.199]$ & \\
\textbf{\texttt{CONTRIBUTING}} \\
1. General contribution instructions     &   & $1.196^{*}$       \\
             &  & $ [1.040; 1.351]$ \\
2. Code style guide    &  & $1.434^{*}$       \\
             &  & $ [1.254; 1.615]$ \\
3. Contributing procedure    &  & $1.270^{*}$       \\
             &  & $ [1.201; 1.338]$ \\
4. Contributor agreement     &  & $1.012^{*}$       \\
             &  & $ [0.848; 1.176]$ \\
5. Build instructions    &  & $1.243^{*}$       \\
             &  & $ [1.123; 1.362]$ \\
\hline
AIC        	& $14648.140$     	&  $2545.991$     	\\
BIC        	& $14711.630$   	&  $2573.417$      	\\
Log Likelihood & $-7314.070$    	& $-1266.996$    	\\
Deviance   	&  $2385.415$   	& $361.955$     	\\
Num. obs.  	& $4226$\phantom{.000}        	& $714$\phantom{.000}         	\\
\hline
\multicolumn{3}{l}{\scriptsize{$^*$ Null hypothesis value outside the confidence interval.}}
\end{tabular}
  \label{tab:topics-outcomes}
\end{table}

\section{Discussion}

\subsection{Project lifecycles and governance}
The analysis we report contributes the first large-scale empirical study of initial documentation in FLOSS projects. Our results show that FLOSS projects introduce both \texttt{README} and \texttt{CONTRIBUTING} files in ways that contrast with the prevailing consensus about the utility of these documents in attracting contributors to a project. \texttt{README} files are published very early on, while \texttt{CONTRIBUTING} files follow (instead of precede) growth in commit activity. Many \texttt{README} files start off short and focus primarily on project usage and installation. Initial versions of \texttt{CONTRIBUTING} files are longer, and tend to provide technical instructions and specify desired contributions. However, \texttt{CONTRIBUTING} documents also deviate from suggestions and do not include characterizations of contributor community. While some features of \texttt{README} and \texttt{CONTRIBUTING} files relate to variations in subsequent activity, such relationships appear tenuous and we find no clear pattern of specific documentation features and subsequent contribution activity.

Both types of files were relatively short: 1658 \texttt{README} files (39\% of our dataset) would take less than 10 seconds to read, while 187 \texttt{CONTRIBUTING} files (26\% of our dataset) are just as short. Such brevity suggests that at initial publication, maintainers either do not view the contents of these files as valuable tools in the development of their project or simply lack the information to populate the documents. Either scenario implies that when these documents are published, they lack content that prospective contributors look for such as thorough project descriptions or comprehensive contributing guidelines \cite{qiu_signals_2019}. 

Publishing a \texttt{README} file at the very beginning of a project's life may be a widespread norm. Platform incentives like GitHub's ``community standards'' checklist may reward the publication of \texttt{README} files as a signifier of project legitimacy, rather than as the community development tools that the platform wants to promote \cite{kerr_folly_1975}. Yet \texttt{README}s created for such reasons may have diminished returns. Projects that publish short, simple \texttt{README} files tend to experience a \textit{steeper} decline in post-publication commit activity than those who publish more longer, more complex files. Further support from practitioner communities and repository platforms (e.g. tools suggesting document content or structure) may be a helpful intervention in provoking more thoughtful and meaningful documentation earlier in project development. 

We also found that projects introduce \texttt{CONTRIBUTING} files in ways that contradict prevailing wisdom. Crucially, the publication of \texttt{CONTRIBUTING} files often came \textit{after} a rise in commit activity. This may imply that publishing a \texttt{CONTRIBUTING} document after an influx of activity reflects a desire to ``get the house in order,'' formalizing already implemented practices. Our topic analysis found that these files emphasized the functional instructions and rules of contributing procedure, rather than addressing dimensions of community or contributor conduct. 


\subsection{Governance as hygiene, governance as catalyst}

The pattern of our findings suggests that FLOSS projects may produce initial \texttt{README} and \texttt{CONTRIBUTING} documentation as a matter of project hygiene rather than as tools for growing a contributor community. On one hand, this might indicate that FLOSS projects are under-utilizing documentation. Future work examining the evolution of documentation files (e.g., common changes in structure, length, content) would offer valuable insight into whether there are opportunities to support earlier-stage writing and development of these files with tooling and suggestions.  On the other, this might imply that the production of FLOSS is not heavily dependent on documentation for initial recruitment of communities of contributors. Deeper investigation into activity preceding and following introduction of documentation, e.g., the burst of attention before \texttt{CONTRIBUTING} files are initially adopted, would help clarify the relationship of these files to contributor recruitment. 


Our results indicate that the projects in our sample did not create \texttt{README} or \texttt{CONTRIBUTING} files in a manner consistent with the common recommendations that such documentation should be designed to draw in future contributions and community engagement. This is not necessarily harmful to project sustainability. For example, projects may not aim to develop contributor communities to begin with. Prior interview studies have identified multiple FLOSS maintainers who do not accept pull requests and generally seek to keep development work confined to themselves \cite{asparouhova_working_2020}. For these maintainers, open source projects are a way of sharing code that they have written, not a way of collaborating with others. An overwhelming research focus on community may misconstrue the actions of project contributors who are disinterested in collaboration. 

However, because the projects in our sample are part of the Debian GNU/Linux distribution, they are widely used and there will be downstream consequences if maintenance decays. 
Our results show that community creation and maintenance may not have been a priority at the time documentation was created; even though, at present, these widely-used projects can benefit from the risk management of active contributor communities \cite{champion_sources_2024, champion_underproduction_2021, walden_impact_2020}. These results illustrate a tension at the heart of FLOSS development: for small projects, community management may be burdensome or counterproductive (especially early in project lifecycles). Nevertheless, the work of community management can become critical for the long term success of the project, especially if it happens to grow to larger scale. 

Project maturity may be an important factor in shaping \texttt{README} and \texttt{CONTRIBUTING} files. Our findings showed that developers initialized files early in a project's life, with little content (sometimes, blank). Universal recommendations of documentation development may lead projects to publish underdeveloped documents in a rote manner, which could result in path dependencies that influence the characteristics of later versions. Notably, recommendations advocating for the creation of \texttt{README} and \texttt{CONTRIBUTING} files rarely specify the exact moment or phase of a project's life where community-building documentation would be most useful. 
Although we focus on two examples of early governance documentation, open source projects make many early decisions---including decisions about language, architecture, license, bug and feature requests, versioning and releases, project ownership, dependencies, vulnerability reporting, and communication flows with end users. Advice on any of these topics may be ignored or even be counterproductive if the goals of project creators and the realities of project growth do not align with the assumptions of those offering advice about best practices. 
Further specificity as well as empirical evaluation are necessary to support best practices for projects of different age and maturity.


\section{Limitations and Future Work}
\label{sec:limitations}

Although the projects we analyze are an accurate snapshot of an important FLOSS ecosystem in current use, the median age of a project in our data set is over twelve years old. 
More recently developed projects or projects created in other ecosystems may introduce \texttt{README} and \texttt{CONTRIBUTING} files differently.
Projects can change version control tools, however our data collection did not include any project history prior to the adoption of git. Moreover, we evaluate commit activity in terms of pure size; the projects in our data set did not have many merge commits. However, previous research has identified approaches to contribution classification \cite{fang_novelty_2024}. Further research should identify the impact that contribution guidelines and additional project governance processes have on different kinds of commit activity. 

Though topic counts were established using term-centric stability analysis, at least one topic generated from our LDA model appears to be ill-defined. Prototypical documents for Topic 7 for the \texttt{README} document data set sometimes lacked topic keywords or semantic themes for the descriptive label of functionality summaries for command-line tools. Moreover, we did not analyze document structure (e.g. headers, outlines) and the possible relationships of those file characteristics to project activity. Given a range of document file types, we did not analyze the use of specific markup languages(e.g. Markdown); the standardized syntax of markup documents may facilitate further insights into document structure. 

Finally, this work makes novel contributions to the understanding of a critical event in FLOSS project life cycles: the initial publication of two kinds of project documents. However, projects continue to evolve as they age, including with the recent popularity of \texttt{GOVERNANCE} documents \cite{chakraborti_we_2023, yan_github_2023} and codes of conduct. Further research is necessary to understand what kinds of decisions are most influential at which points in a project's life cycle.

\section{Conclusion}
We found that projects largely introduce \texttt{README} files at the beginning of their life cycle, but create \texttt{CONTRIBUTING} documents later, following increased project activity. Moreover, we found that documents are largely focused on the functional processes of usage and contribution, not on the community development for which these files are often recommended. 
Efforts to promote FLOSS project sustainability often emphasize the development of communities. Yet, recommendations that assume that project founders want to create communities (and that process documents can catalyze this) may fail to address the realities of early-lifecycle projects. 
Too much emphasis on community-building could run counter to project founders' and contributors' diverse motives, generating a sense of bureaucracy, complexity, and weighty commitment. This, in turn, could discourage participation. At the same time, our findings suggest that more substantive project documentation may be associated with subsequent project growth. Recommendations for best practices and governance should evaluate these possibilities directly. We hope that our data and findings will bring more research attention to how projects use documentation and how these files impact project health.


\section*{Acknowledgment}

This work is indebted to the volunteers producing FLOSS who have made their work available for inspection. We also gratefully acknowledge support from the Ford/Sloan Digital Infrastructure Initiative (Sloan Award 2018-113560) and the National Science Foundation (Grant IIS-2045055).  This work was conducted using the Hyak supercomputer at the University of Washington as well as research computing resources at Northwestern University.



%
\bibliographystyle{IEEEtran}
\bibliography{debgov_bib}

\end{document}